\documentclass{appolb}
\usepackage{epsfig}

\begin{document}

\title{SEARCH FOR FINGERPRINTS OF TETRAHEDRAL SYMMETRY 
       IN $^{156}$Gd\thanks{Presented at the Zakopane Conference on Nuclear
       Physics, September 1-7, 2008}}
       
\author{
Q.T. Doan$^{1}$, D. Curien$^{2}$, O. St\'{e}zowski$^{1}$, J.~Dudek$^{2}$, K.~Mazurek$^{5}$, A.~G\'o\'zd\'z$^{3}$, J.~Piot$^{2}$, G.~Duch\^{e}ne$^{2}$, B. Gall$^{2}$, H.~Molique$^{2}$, M.~Richet$^{2}$, P.~Medina$^{2}$,D.~Guinet$^{1}$, N.~Redon$^{1}$, Ch.~Schmitt$^{1}$, P.~Jones$^{4}$, P.~Peura$^{4}$, S.~Ketelhut$^{4}$, M.~Nyman$^{4}$, U.~Jakobsson$^{4}$, P.T. Greenlees$^{4}$, R.~Julin$^{4}$, S. Juutinen$^{4}$, P. Rahkila$^{4}$, A.~Maj$^{5}$, K.~Zuber$^{5}$, P.~Bednarczyk$^{5}$, N.~Schunck$^{6}$, J.~Dobaczewski$^{4,7}$, A.~Astier$^{8}$, I.~Deloncle$^{8}$, D.~Verney$^{9}$, G.~de~Angelis$^{10}$ and J.~Gerl$^{11}$
\address{
$^{1}$IPN Lyon, Universit\'{e} Lyon 1, IN2P3-CNRS, F-69622 Villeurbanne, France\\
$^{2}$IPHC-DRS, ULP, IN2P3-CNRS, 
F-67037 Strasbourg, France\\
$^{3}$Dept. of Math. Phys., Uni. Marii Curie-Sk\l{}odowskiej,
 PL-20031 Lublin, Poland \\
$^{4}$Department of Physics, University of Jyv\"askyl\"a, FI-40014 Jyv\"askyl\"a, Finland\\
$^{5}$Niewodnicza\'nski Institute of Nuclear Physics PAN, PL-31-342 Krak\'ow, Poland\\
$^{6}$Physics Division, ORNL, Oak Ridge, TN-37831, USA\\
$^{7}$Institute of Theoretical Physics, Warsaw University, PL-00-681 Warsaw, Poland\\
$^{8}$CSNSM, IN2P3-CNRS, F-91405 Orsay Campus, France\\
$^{9}$IPN Orsay, IN2P3-CNRS, F-91406 Orsay Cedex, France\\
$^{10}$INFN, Laboratori Nazionali di Legnaro, I-35020 Legnaro, Italy\\
$^{11}$GSI, D-64291 Darmstadt, Germany}}
\maketitle

%%%%%%%%%%%%%%%%%%%%%%%%%%%%%%%%%%%%%%%%%%%%%%%%%%%%%%%%%%%%%%%%%%%%%%%%%%%%%%%%
%%%%%%%%%%%%%%%%%%%%%%%%%%%%%%%%%%%%%%%%%%%%%%%%%%%%%%%%%%%%%%%%%%%%%%%%%%%%%%%%

\begin{abstract}

Theoretical predictions suggest the presence of tetrahedral symmetry as an
explanation for the vanishing intra-band E2 transitions at the bottom of the
odd-spin negative-parity band in $^{156}$Gd. The present study reports on
experiment performed to address this phenomenon. It allowed to remove certain
ambiguouities related to the intra-band E2 transitions in the negative-parity
bands, to determine the new inter-band transitions and reduced probability
ratios $B$(E2)/$B$(E1) and, for the first time, to determine the experimental
uncertainties related to the latter obervable.

\end{abstract}


%%%%%%%%%%%%%%%%%%%%%%%%%%%%%%%%%%%%%%%%%%%%%%%%%%%%%%%%%%%%%%%%%%%%%%%%%%%%%%%%
%%%%%%%%%%%%%%%%%%%%%%%%%%%%%%%%%%%%%%%%%%%%%%%%%%%%%%%%%%%%%%%%%%%%%%%%%%%%%%%%

\section{Introduction}

It has been suggested on the basis of the realistic nuclear mean-field
calculations, Ref.~\cite{Li94}, that there should exist atomic nuclei whose
shapes are tetrahedral-symmetric. Theoretical calculations, Ref.~\cite{Du06} and
references therein, suggest that to a first approximation, the nuclei whose
shapes are characterized by the exact tetrahedral symmetry have vanishing
multipole moments except for the $Q_{3\pm2}$ one, the next order multipoles
allowed by the tetrahedral symmetry are $Q_{7\pm2}$ and $Q_{7\pm6}$ and such
contributions are expected to be very small if not totally neglegible in the
nucleus studied. Thus, unlike in rotational bands of quadrupole-deformed nuclei
where the E2 transitions dominate, in tetrahedral bands the E2 transitions are
predicted to vanish or to be very weak, because the quadrupole moments go to
zero when the tetrahedral symmetry becomes exact. According to ENSDF,
Ref.~\cite{ensdf}, the nucleus $^{156}$Gd  has been studied in over 15 different
excitation modes with varying target-beam combinations, beam energies, and
detection systems. Although a regular sequence of odd-spin negative-parity
states has been established down to $I^\pi=3^-$, the intra-band E2 transitions
below the $I^\pi=9^{-}$ state have never been seen. The energies of the
corresponding states have been measured exclusively through the inter-band E1
transitions to the ground-state band. Such a behavior is expected to be a
consequence of tetrahedral symmetry \cite{Du06}. Already in the early eighties,
Konijn and co-workers, Ref.~\cite{Ko81}, carried out an experiment using an
$\alpha$-particle beam and measured the ratios of the reduced transition
strengths, $B$(E2)/$B$(E1), for two negative-parity bands in $^{156}$Gd -- at
that time interpreted as octupole vibrational bands. The $B$(E2)/$B$(E1) ratios
were found to be about a factor 50 lower in the odd-spin, as compared to those
in the even-spin negative-parity bands. More recently, Sugawara,
Ref.~\cite{Sug01}, measured the branching ratios of these two negative-parity
bands by using the reaction $^{150}$Nd($^{13}$C,$\alpha$3n). In the case of the
odd-spin negative-parity bands, a minimum in the $B$(E2)/$B$(E1) ratios at
intermediate spins was reported and some upper limits of branching ratios at low
spins were measured. These measurements have been carried out at best by using
the $\gamma-\gamma$ coincidences with a population of $^{156}$Gd that may not
have been enough to observe the low-intensity transitions. The main goal of this
experiment was to search for the E2 transitions forbidden by the tetrahedral
symmetry with high statistics, to determine the $B$(E2)/$B$(E1) ratios, and to
search for any signs of cross-feeding involving the odd-spin negative-parity
band.

%%%%%%%%%%%%%%%%%%%%%%%%%%%%%%%%%%%%%%%%%%%%%%%%%%%%%%%%%%%%%%%%%%%%%%%%%%%%%%%%
%%%%%%%%%%%%%%%%%%%%%%%%%%%%%%%%%%%%%%%%%%%%%%%%%%%%%%%%%%%%%%%%%%%%%%%%%%%%%%%%

\section{Experiment}

The nucleus $^{156}$Gd was produced by using the fusion-evaporation reaction
$^{154}$Sm($\alpha$,2n) and then studied by using the JUROGAM $\gamma$-ray
detector, Ref.~\cite{jurogam}, at Jyv\"{a}skyl\"{a}. The optimal bombarding
energy (27~MeV)  was deduced from the excitation function measured for this
reaction at the Orsay Tandem during a pilot experiment. This bombarding energy
enabled us to optimize the population at low and medium spins in $^{156}$Gd and
to minimize the contaminations from other channels (\eg mainly $^{155}$Gd) below
8\%. In this experiment, 43 Anti-Compton suppressed HP-Ge detectors were used,
giving a total photopeak efficiency of 4.2\%. We used self-supporting, 99.2\%
enriched, $^{154}$Sm targets with a thickness of 2~mg/cm$^{2}$.  The acquisition
was performed by both analogue and digital system in triggerless mode. The TNT2
digital acquisition cards from the IPHC, Strasbourg, were used to record data 
from prompt gamma-ray emissions from the Germanium detectors. The digital
acquisition allows a higher count-rate (up to 100~kHz) due to shorter deadtime
\cite{Arno05}.  The digitization of the ADC pulse via the Jordanov algorithm
\cite{Jord94} provides a  stable energy measurement and fast baseline
restoration. These features provide access  to a wider range of beam intensities
and therefore to phenomena with lower cross sections. At a similar count-rate,
the digital acquisition records 36\% more statistics  than the analogue system
and shows a better linearity in energy, specifically under 300~keV. In our
study, a total of $228 \times 10^{6}$ $\gamma\gamma\gamma$ coincidence-events
have been collected (i.e. pure unfolded coincidences after Compton-suppresion).

%%%%%%%%%%%%%%%%%%%%%%%%%%%%%%%%%%%%%%%%%%%%%%%%%%%%%%%%%%%%%%%%%%%%%%%%%%%%%%%%
%%%%%%%%%%%%%%%%%%%%%%%%%%%%%%%%%%%%%%%%%%%%%%%%%%%%%%%%%%%%%%%%%%%%%%%%%%%%%%%%

\section{Results}

\begin{figure}[htb!]
  \begin{center}
  \includegraphics[width=0.85\textwidth]{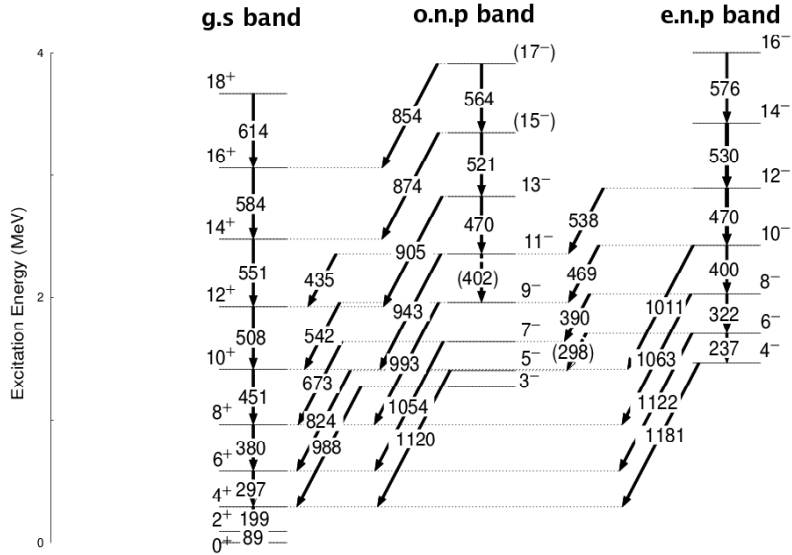}
  \caption{Partial decay scheme of ${}^{156}$Gd showing the ground-state band,
           odd- and and even-spin negative parity bands; the newly
           established interconnecting transitions are shown (see the text).}
                                                            \label{fig:levelGd}
  \end{center}
\end{figure}

A partial level scheme of $^{156}$Gd, established in this work, is displayed in
Fig.~\ref{fig:levelGd}. For the odd-spin negative-parity band we confirm that
the E2 transitions vanish below the $I^\pi=9^{-}$ state. The intensity of the
$11^{-} \rightarrow 9^{-}$ transition is very weak and could not be firmly
established. In fact, this transition is part of a doublet (400-402 keV) in
coincidences with another doublet at 470 keV both present in the odd-spin band
and the even spin band. Therefore gating on the 470 keV line to extract the 402
keV intensity would bring in any case residual contamination from the 400 keV
line. 
The E1 transitions de-exciting the $I^\pi=1^{-}$ state and the E2 transition
connecting the $4^{-} \rightarrow 2^{-}$ (reported in previous experiments) from
the even-spin band cannot be confirmed by our results. However,
$\gamma\gamma\gamma$ coincidences allowed us to clarify a number of
uncertainties caused by the presence of doublet- and even triplet-lines in the
spectrum of this nucleus. Moreover, we were able to examine the transitions in
the medium spin range and firmly establish new inter-band transitions with
$E_\gamma$ of 538, 469 and 390~keV. Angular distributions will be analysed in
the near future to determine the character (stretched M1 or non-stretched E2) of
these transitions. Table \ref{tab:Bratios} shows some preliminary
$B$(E2)/$B$(E1) ratios that we have found, compared to the results of the
previous studies of Refs.~\cite{Ko81,Sug01}. For the $15^{-}$ and $13^{-}$
states of the odd-spin negative-parity band, the transition strength ratios are
of the same order of magnitude as previously reported, while for higher spin
states they could not be determined because of the cut-off in angular momentum
due to the use of the $\alpha$-beam.
\begin{table}[!h]
\begin{center}
\begin{tabular}{ccccccc}
\hline \hline
{\small $I^\pi$} & \multicolumn{2}{c}{{\small \quad $B$(E2)/$B$(E1)}} & \hspace{0.6cm} &
{\small $I^\pi$} & \multicolumn{2}{c}{{\small \quad $B$(E2)/$B$(E1)}} \\
& {\small (a)\quad} & {\small (b)$\;\;$} & & & {\small (a)\quad} & {\small (b)} \\
\hline
{\small $17^-$} & {\small ---} & {\small 16(3)} && {\small $12^-$} & {\small ---} & \\
{\small $15^-$} & {\small 4.5(1.0)} & {\small 6(2)} && {\small $10^-$} & {\small 640(100)} & {\small 240} \\
{\small $13^-$} & {\small 5.5(0.6)}	& {\small 7(2)} && {\small $8^-$} & {\small 330(10)} & {\small 700} \\
{\small $11^-$} & {\small $<$9(-2)} & {\small 15(7)} && {\small $6^-$}	& {\small 210(15)} &	{\small 350}\\
{\small $9^-$}	& {\small $<$26(-5)}	&& & {\small $4^-$} & {\small ---} & \\
{\small $7^-$}	& {\small $<$92(-11)} && & & \\
{\small $5^-$}	& {\small ---} && & & \\
\hline \hline
\end{tabular}
\end{center}
\caption{Branching ratios $B$(E2)/$B$(E1) in units $10^{\,6}$fm$^2$. 
        (a) Established in the present work - in comparison with: 
        (b) Previous results from Refs.~\cite{Ko81,Sug01}.}
                                                           \label{tab:Bratios}
\end{table}
Only upper limits are established for the lowest spins, however this information
represents already a progress since no earlier publication quotes any estimates
for the $(9^-\to7^-)$ and $(7^-\to5^-)$ transitions. For the even-spin
negative-parity band the B(E2)/B(E1) ratios decrease with decreasing spin and
are up to two orders of magnitude higher than those of the odd-spin
negative-parity band.

%%%%%%%%%%%%%%%%%%%%%%%%%%%%%%%%%%%%%%%%%%%%%%%%%%%%%%%%%%%%%%%%%%%%%%%%%%%%%%%%
%%%%%%%%%%%%%%%%%%%%%%%%%%%%%%%%%%%%%%%%%%%%%%%%%%%%%%%%%%%%%%%%%%%%%%%%%%%%%%%%

\section{Conclusions and Discussion}

The vanishing of intra-band E2 transitions, supporting the tetrahedral symmetry
interpretation at the bottom of the odd-spin negative-parity band, has been
confirmed along with the two-orders-of-magnitude differences in the
$B$(E2)/$B$(E1) branching ratios of two negative-parity bands. Thanks to the
$\gamma\gamma\gamma$ coincidences, we have established new inter-band
transitions. As it is known from general considerations, cf.\ Ref.~\cite{BMII},
a $K^\pi=0^-$ band must not have even spins and it follows that the even-spin
negative-parity band discussed here must not be interpreted as $0^-$, in
contrast to some first claims in the past. We were not able to establish the
$4^-\to2^-$ transition which most likely signifies that it is very weak or
non-existent. Therefore, it will be even more important to find-out whether the
$\Delta I=1$ transitions connecting the even- and odd-spin negative-parity bands
have the $M1$-character, which could suggest the presence of high-$K$ components
in the underlying band-heads. 

Let us emphasize at this point that the tetrahedral configurations, as predicted
by theory, are markedly non-axial and, therefore, are expected to strongly mix
components of wave-functions with various quantum numbers $K$: the strongest
component associated with the geometry of shapes based on the $Y_{3+2}+Y_{3-2}$
spherical harmonics should be $K=2$. Theoretical calculations based on the
generalised collective rotor Hamiltonian that includes terms of the third order
in angular momentum\footnote{Hamiltonians of order higher than two in terms of
the angular momentum operators are commonly used in molecular physics to
describe the geometrical molecular symmetries.} indicate that the structure of
the wave-function of the $1^-$ state is exceptional since, in contrast to states
with $I\geq2$, it {\em must not manifest} the tetrahedral symmetry. In other
words, for $I\leq1$ the tetrahedral symmetry is excluded; actually $1^-$ state
wave-function manifest an axial symmetry. Consequently, the role of the $1^-$
state, often treated as a member of the (expected to be) the tetrahedral band,
is special in that even if connected  to the $3^-$ state via an E2 transition,
in principle possible due to an expected to be strong a K-mixing, its underlying
symmetry must not be tetrahedral. Our experiement, similarly to the preceding
ones, gives no sign of the $3^-\to1^-$ transition neither, what signifies that
the corresponding E2 transition, if exists, must be very weak.

%%%%%%%%%%%%%%%%%%%%%%%%%%%%%%%%%%%%%%%%%%%%%%%%%%%%%%%%%%%%%%%%%%%%%%%%%%%%%%%%
%%%%%%%%%%%%%%%%%%%%%%%%%%%%%%%%%%%%%%%%%%%%%%%%%%%%%%%%%%%%%%%%%%%%%%%%%%%%%%%%

\section*{Acknowlegements}

This work has benefited from the use of TNT2-D cards, developed and financed by CNRS/IN2P3 for the GABRIELA project. It was partially supported by the EU through the EURONS project under
contract No.~RII3 CT 2004~506065, by the collaboration Tetra-Nuc through the
IN2P3, France, by the Academy of Finland and University of Jyv\"askyl\"a within
the FIDIPRO programme, and by the Polish Ministry of Science under Contract
No.~N N202 328234. We thank the Direction of IPNO and participating members of
IPNO and CSNSM for helping us to launch the first TetraNuc pilot experiment at
ALTO. Warm thanks are addressed to both IPNO and JYFL accelerator staffs.

\end{document}